\documentclass[letterpaper]{article} % DO NOT CHANGE THIS
\usepackage[]{aaai25}  % DO NOT CHANGE THIS
\usepackage{times}  % DO NOT CHANGE THIS
\usepackage{helvet}  % DO NOT CHANGE THIS
\usepackage{courier}  % DO NOT CHANGE THIS
\usepackage[hyphens]{url}  % DO NOT CHANGE THIS
\usepackage{graphicx} % DO NOT CHANGE THIS
\usepackage{amsmath}  
\usepackage{amssymb}  
\usepackage{natbib}   
\usepackage{caption}  
\usepackage{booktabs}
\usepackage{pifont}
\usepackage{colortbl}
\usepackage{xcolor}

\usepackage{algorithm}
\usepackage{algpseudocode}

\usepackage{float}
\usepackage{makecell}
\newcommand{\cmark}{\ding{51}}  % 对号 ✓
\newcommand{\xmark}{\ding{55}}  % 叉号 ✗
\usepackage{newfloat}
\usepackage{listings}
\DeclareCaptionStyle{ruled}{labelfont=normalfont,labelsep=colon,strut=off} % DO NOT CHANGE THIS
\floatstyle{ruled}
\newfloat{listing}{tb}{lst}{}
\floatname{listing}{Listing}
\title{SlimRAG: Retrieval without Graphs via Entity-Aware Context Selection}
\author{
  Jiale Zhang\textsuperscript{1,3,4*},
  Jiaxiang Chen\textsuperscript{1,2*} 
  Zhucong Li\textsuperscript{1,2}\textsuperscript{\dag}, 
  Jie Ding\textsuperscript{3,4}, 
  Kui Zhao\textsuperscript{3,4}, 
  Zenglin Xu\textsuperscript{2}, 
  Xin Pang\textsuperscript{1}, 
  Yinghui Xu\textsuperscript{2}
}

\affiliations {
  \textsuperscript{1}Continue-AI \\
  \textsuperscript{2}Fudan University \\
  \textsuperscript{3}University of Chinese Academy of Sciences \\
  \textsuperscript{4}Shenyang Institute of Computing Technology, Chinese Academy of Sciences \\
  \texttt{zhangjiale23@mails.ucas.ac.cn, \{jiaxiangchen23,zcli22\}@m.fudan.edu.cn}
}
\makeatletter
\def\copyright@text{%
  \hspace*{1.5em}%
  \textsuperscript{*}Equal contribution. \quad
  \textsuperscript{\dag}Corresponding author.\\
  \hspace*{1.5em}%
  This work is in progress and will be continuously updated.
}

\begin{document}
\maketitle

% Abstract

\begin{abstract}
Retrieval-Augmented Generation (RAG) enhances language models by incorporating external knowledge at inference time. However, graph-based RAG systems often suffer from structural overhead and imprecise retrieval: they require costly pipelines for entity linking and relation extraction, yet frequently return subgraphs filled with loosely related or tangential content. This stems from a fundamental flaw—\emph{semantic similarity} does not imply \emph{semantic relevance}.
We introduce \textbf{SlimRAG}, a lightweight framework for retrieval without graphs. SlimRAG replaces structure-heavy components with a simple yet effective entity-aware mechanism. At indexing time, it constructs a compact entity-to-chunk table based on semantic embeddings. At query time, it identifies salient entities, retrieves and scores associated chunks, and assembles a concise, contextually relevant input—without graph traversal or edge construction.
To quantify retrieval efficiency, we propose \textbf{Relative Index Token Utilization (RITU)}, a metric measuring the compactness of retrieved content. Experiments across multiple QA benchmarks show that SlimRAG outperforms strong flat and graph-based baselines in accuracy while reducing index size and RITU (e.g., 16.31 vs. 56+), highlighting the value of structure-free, entity-centric context selection. The code will be released soon\footnote{\url{https://github.com/continue-ai-company/SlimRAG}}.

\end{abstract}

% Introduction

\section{Introduction}
\begin{figure}[htb!]
  \centering
  \includegraphics[width=1.0\linewidth]{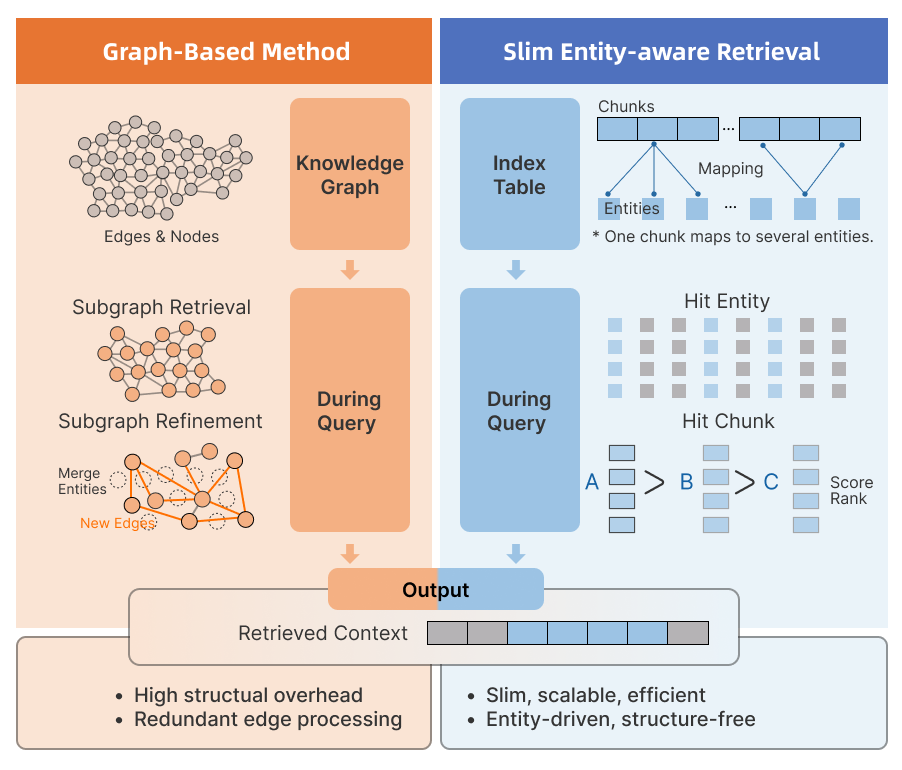}
\caption{
Comparison between graph-based methods and our SlimRAG framework.
Graph-based approaches suffer from high structural overhead due to graph construction and costly subgraph refinement, particularly in generating new relational edges.
SlimRAG provides a structure-free alternative by directly retrieving and composing context through lightweight entity-to-chunk indexing.
}

\label{fig:comparison-slimrag}
\end{figure}

Retrieval-Augmented Generation (RAG) has become a cornerstone of knowledge-intensive NLP applications such as question answering, scientific summarization, and decision support~\cite{lewisRetrievalaugmentedGenerationKnowledgeintensive2020, huangSurveyRetrievalAugmentedText2024}. By retrieving and incorporating external knowledge at inference time, RAG enables language models to generate factually grounded and context-aware responses that go beyond what is stored in model parameters.

However, a fundamental bottleneck persists: \emph{retrieval redundancy}. Standard RAG pipelines typically adopt \textit{flat indexing}, segmenting documents into fixed-size passages or entity descriptions, embedding them independently, and retrieving top-scoring matches by vector similarity~\cite{guoLightRAGSimpleFast2024, liEfficientDynamicClusteringBased2025}. This often results in large, overlapping, and poorly scoped retrieval sets that contain significant irrelevant or marginal content~\cite{wuHowEasilyIrrelevant2024a}. To address this, a growing body of work attempts to introduce structural knowledge through \textit{graph-based indexing}—linking entities and retrieving subgraphs~\cite{edgeLocalGlobalGraph2024, somanBiomedicalKnowledgeGraphenhanced2023, chenSACKGExploitingLarge2024}.

Graph-based methods face fundamental challenges in dynamic or large-scale retrieval settings. They rely on accurate named entity recognition, linking, and relation extraction—components that are fragile and error-prone in open-domain contexts. As corpora expand, graphs become increasingly difficult to maintain and update. Even after retrieving subgraphs, systems often perform costly refinement steps such as entity merging and new edge generation. These operations add computational overhead, and the resulting subgraphs typically require linearization before they can be processed by language models~\cite{xuGenerateonGraphTreatLLM2024}.

Crucially, structural complexity does not guarantee retrieval quality. Many graph edges capture weak or tangential relationships that introduce noise rather than aid reasoning. This is compounded by a deeper mismatch: \textbf{semantic similarity does not imply semantic relevance}. Graphs constructed from co-occurrence patterns or embedding proximity may retrieve topically related content, but not necessarily information aligned with the query intent. These issues highlight the limitations of structure-heavy pipelines in achieving precise, context-sensitive retrieval.

To this end, we propose \textbf{SlimRAG}, a graph-free, entity-aware retrieval framework. At indexing time, SlimRAG constructs a compact entity-to-chunk mapping that organizes knowledge without relying on structural edges. At query time, it identifies salient entities from the question, retrieves associated chunks, applies deduplication and ranking, and produces an ordered context for generation—entirely without graph traversal or edge manipulation. This architecture is inherently scalable and robust, avoiding the complexity and brittleness of structural reasoning.

To quantify retrieval efficiency and index compactness, we introduce \textbf{Relative Index Token Utilization (RITU)}, a new metric that measures the proportion of unique corpus tokens retained in the index. Experiments across multiple QA benchmarks show that SlimRAG matches or outperforms strong flat and graph-based baselines in answer accuracy, while achieving dramatically smaller index size, lower retrieval redundancy, and faster query-time response. These findings highlight the value of lightweight, relevance-driven design in building the next generation of scalable RAG systems.

\noindent \textbf{Our main contributions are:}
\begin{itemize}
\item We identify structural over-specification—driven by the mistaken equivalence of \emph{semantic similarity} and \emph{semantic relevance}—as a key source of redundancy in RAG systems, and advocate a graph-free, entity-centric alternative.

\item We propose \textbf{SlimRAG}, a graph-free, entity-aware retrieval framework that decouples entity recognition from relation modeling, enabling efficient and scalable context selection without explicit structural reasoning.

\item We introduce \textbf{Relative Index Token Utilization (RITU)}, a novel metric that quantifies index compactness by measuring the proportion of retained corpus tokens. SlimRAG achieves strong retrieval performance with significantly reduced index size and redundancy.

\end{itemize}

\section{Preliminaries}
\label{sec:preliminaries}

\subsection{Semantic Similarity vs. Semantic Relevance}

\begin{figure}[t]
  \centering
  \includegraphics[width=1.0\linewidth]{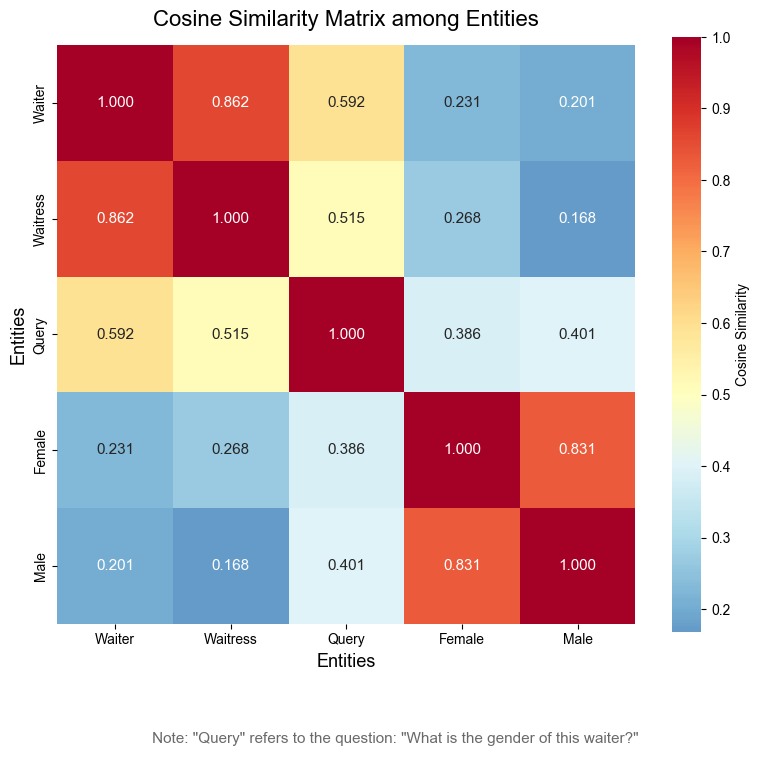}
  \caption{
    \textbf{Semantic similarity may mislead relevance-based retrieval.} Cosine similarity heatmap between the query \textit{``What is the gender of this waiter?''} and various related entities. Although \textit{Waiter} and \textit{Waitress} show the highest similarity, they are not useful for answering the query. More relevant entities like \textit{Male} are semantically farther, underscoring the need to distinguish similarity from relevance during retrieval.
  }
  \label{fig:heatmap_matrix}
\end{figure}

As illustrated in Figure~\ref{fig:heatmap_matrix}, semantic similarity in embedding space often fails to capture the task-specific relevance needed for accurate retrieval. Consider the query: \textit{``What is the gender of this waiter?''}. One might expect entities like \textit{Male} or \textit{Female} to be highly relevant. However, as shown in the cosine similarity heatmap, \textit{Waiter} and \textit{Waitress} share the highest similarity—despite \textit{Waitress} being irrelevant or even misleading for answering the query.

This observation motivates a crucial distinction between two concepts:

\textbf{Semantic similarity} refers to the general, query-independent closeness between two items in a global embedding space trained over broad corpora. It captures overall contextual or lexical resemblance—e.g., \textit{Waiter} and \textit{Waitress} appear in similar settings, leading to a high cosine similarity.

\textbf{Semantic relevance}, by contrast, is query-sensitive. It measures how pertinent an entity or passage is in addressing a specific information need. In our example, \textit{Male} is more relevant to the question about gender, even though its embedding similarity to \textit{Waiter} may be lower.

Neglecting this distinction can lead retrieval systems to prioritize surface-level proximity over actual informativeness—retrieving content that is similar but not helpful for the user’s intent.

\subsection{Design Implications: From Conceptual Distinction to System Design}

The conceptual separation between \textbf{semantic similarity} and \textbf{semantic relevance} informs the modular design of SlimRAG, enabling a structured yet lightweight retrieval pipeline that avoids the overhead of explicit graph construction.

\begin{itemize}
    \item \textbf{Similarity for Structure:} In the indexing phase, entities extracted from the corpus are embedded into a global semantic space and grouped based on similarity. This enables compact indexing, efficient access, and reuse—without modeling complex inter-entity relations or graph edges.
    
    \item \textbf{Relevance for Retrieval:} During query-time inference, we prioritize context that is not just globally similar but specifically relevant to the query intent. Each candidate chunk is scored based on both embedding similarity to the query and the number of matched entities it contains.
\end{itemize}

This separation of concerns—using similarity to organize knowledge and relevance to guide retrieval—allows SlimRAG to achieve precise, task-aware context selection without relying on graph-based architectures. 

As illustrated in Figure~\ref{fig:heatmap_matrix}, treating high similarity as high relevance can lead to misleading retrieval. SlimRAG avoids this pitfall by anchoring retrieval in entity-level relevance, ensuring alignment with the user's actual information need.

\section{Method}
\label{sec:methodology}
\begin{figure*}[t]
  \centering
  \includegraphics[width=1.0\linewidth]{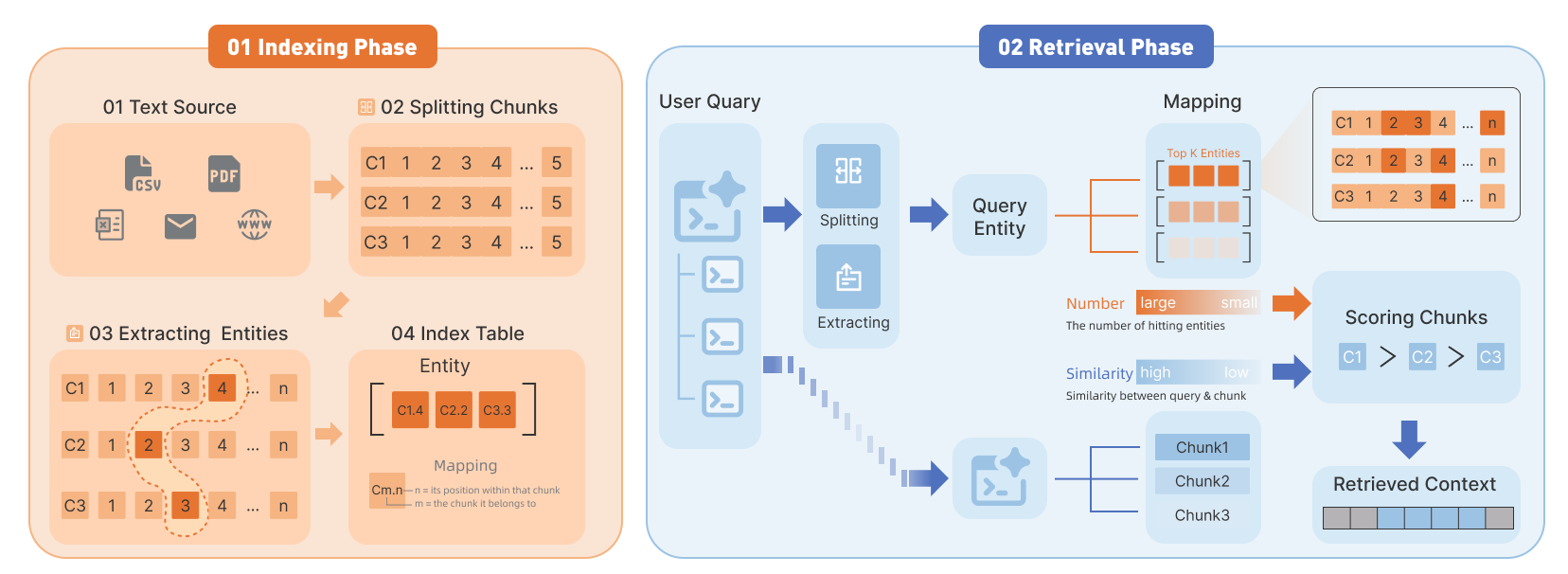}
\caption{
\textbf{Overview of the SlimRAG pipeline.} The framework consists of two main stages: \textit{Indexing Phase} and \textit{Retrieval Phase}. In the indexing stage (left), raw documents from heterogeneous sources (e.g., CSV, PDF, web) are segmented into chunks, from which salient entities are extracted. These entities are then organized into a lightweight inverted index that maps each entity to the set of chunks in which it appears. During retrieval (right), given a user query, SlimRAG extracts entities from the query, performs semantic matching against the indexed entities, and retrieves the most relevant chunks based on both entity overlap and semantic similarity. The selected chunks are then ranked and assembled as retrieved context for downstream tasks such as reasoning or generation.
}

  \label{fig:slimrag-pipeline}
\end{figure*}

This paper introduces \textbf{SlimRAG}, a lightweight and entity-aware retrieval framework designed to extract contextually relevant chunks from a large unstructured corpus $\mathcal{C}$. Given a natural language query $q$, the goal is to assemble a concise and informative context $Ctx$ that supports downstream applications such as reasoning, question answering, and summarization.

SlimRAG consists of two key stages: (1) an \textbf{Indexing Phase}, which builds a compact inverted index over extracted entities, and (2) a \textbf{Retrieval Phase}, which selects and ranks chunks based on their semantic relevance to the query. The following sections detail the design and implementation of each component.

\subsection{Indexing Phase: Entity-Aware Preprocessing for Efficient Retrieval}
\label{ssec:indexing}

The \textbf{Indexing Phase} transforms the raw corpus $\mathcal{C}$ into a compact structure that enables efficient query-time retrieval. Rather than relying on complex graph structures, SlimRAG adopts a lightweight \textbf{entity-aware inverted indexing} strategy based on chunk-level granularity and semantic matching.

\paragraph{Chunk Segmentation and Entity Extraction.} 
We begin with a collection of raw text chunks $\mathcal{C} = \{C_1, C_2, \dots, C_N\}$ obtained via preprocessing and segmentation (e.g., HTML removal, normalization, and sentence grouping). Each chunk $C_i$ is treated as a coherent unit of content.

For each chunk $C_i \in \mathcal{C}$, we extract a set of normalized entities using a prompt-based function:
\[
E_i = \text{ExtractEntities}(C_i)
\]
This function identifies salient entities—including aliases and coreferent mentions—so that $E_i$ reflects the main semantic referents within the chunk. The sequential order of chunks is retained to support coherence in later stages of context reconstruction.

\paragraph{Index and Inverted Map Construction.} 
We construct a global \textbf{Entity Index} $E_I$ by aggregating all unique entities extracted from the corpus:
\[
E_I = \bigcup_{k=1}^{M} \bigcup_{j=1}^{N_k} E_{k,j}
\]
In parallel, we build an \textbf{Entity-to-Chunk Inverted Map} $M_{IC}$, which maps each entity $e \in E_I$ to the set of chunk identifiers in which it appears:
\[
M_{IC}(e) = \{ (k, j) \mid e \in E_{k,j} \}
\]
This structure enables efficient retrieval of all chunks associated with a given entity, supporting fast and sparse access without needing to traverse or maintain edge relations. Notably, $M_{IC}$ naturally supports aliasing and overlapping mentions, as the same chunk can be indexed under multiple entities. This indexing layer forms the backbone of SlimRAG's structure-free retrieval pipeline.

\paragraph{Comparison and Scalability.} 
Unlike systems such as GraphRAG or LightRAG that rely on edge-heavy knowledge graphs, SlimRAG avoids explicit graph construction, coreference resolution, or costly disambiguation steps. Instead, it performs direct entity-level matching using a lightweight table structure. This design not only eliminates the overhead of edge generation and maintenance, but also makes the system inherently modular and adaptable to incremental updates. By operating over chunk-level granularity, SlimRAG significantly reduces memory and computation costs during indexing—making it scalable to large, continuously growing corpora.

\begin{algorithm}[t]
\caption{SlimRAG: Entity-Guided Context Retrieval}
\label{alg:slimrag}
\small
\begin{algorithmic}[1]
\Require Chunk collection $\mathcal{C} = \{C_1, C_2, \dots, C_M\}$, query $q$, encoder $\text{emb}$, similarity function $\phi$
\State \textbf{// Indexing Phase}
\State Initialize entity set $E_I \gets \emptyset$, entity-to-chunks map $M_{EC} \gets \{\}$
\ForAll{chunk $C_k \in \mathcal{C}$}
    \State $E_k \gets \text{ExtractEntities}(C_k)$
    \State $E_I \gets E_I \cup E_k$
    \ForAll{entity $e \in E_k$}
        \State $M_{EC}(e) \gets M_{EC}(e) \cup \{k\}$
    \EndFor
\EndFor
\State \textbf{// Retrieval Phase}
\State $Q_{\text{sub}} \gets \text{DecomposeQuery}(q)$ \Comment{Using LLM}
\State $E_Q \gets \text{ExtractEntities}(Q_{\text{sub}})$
\State $W_e \gets \text{ComputeEntityWeights}(E_Q, Q_{\text{sub}})$
\State $E_H \gets \emptyset$ \Comment{Hit entities}
\ForAll{entity $e_q \in E_Q$}
    \State $E_q^{(K)} \gets \text{RetrieveTopK}(e_q, E_I, \phi)$
    \State $E_H \gets E_H \cup E_q^{(K)}$
\EndFor
\State $C_H \gets \bigcup_{e \in E_H} M_{EC}(e)$ \Comment{Hit chunks}
\ForAll{chunk $C_k \in C_H$}
    \State $\text{count}_k \gets$ Number of hit entities referring to chunk $k$
    \State $\phi_q \gets \phi(\text{emb}(C_k), \text{emb}(q))$
    \State $\text{score}_k \gets \phi_q \cdot \text{count}_k$
\EndFor
\State Sort chunks by score and select top-K: $C_{top} \gets \text{TopK}(C_H)$
\State Merge texts within token limit: $\text{Context} \gets \text{MergeTexts}(C_{top})$
\State \Return $\text{Context}$
\end{algorithmic}
\end{algorithm}

\subsection{Retrieval Phase: Entity-Guided Context Construction}
\label{ssec:retrieval}
The \textbf{Retrieval Phase} constructs a semantically relevant \textbf{Context} $Ctx$ for a given user query $q$ by leveraging the entity-aware index generated in the Indexing Phase. As outlined in Algorithm~\ref{alg:slimrag}, the retrieval process follows multiple steps that balance semantic similarity with entity-driven relevance.

\paragraph{Query Decomposition and Entity Extraction.}
Given a query $q$, we first decompose it into multiple sub-queries using a large language model:
\[
Q_{\text{sub}} = \text{DecomposeQuery}(q)
\]
This decomposition ensures that complex information needs are broken down into simpler, more focused components. From these sub-queries, we extract the semantic entities:
\[
E_Q = \text{ExtractEntities}(Q_{\text{sub}})
\]
We then compute entity weights $W_e$ based on their frequency across the sub-queries, reflecting their relative importance to the overall query intent.

\paragraph{Entity Matching via Semantic Search.}
To connect the query with the corpus-level entity set $E_I$, we perform semantic search for each entity in $E_Q$. For each query entity $e_q \in E_Q$, we retrieve the top-$K$ most similar corpus entities based on embedding similarity:
\[
E_q^{(K)} = \text{RetrieveTopK}(e_q, E_I, \phi)
\]
where $\phi(\text{emb}(e_q), \text{emb}(e_i))$ represents the similarity function between the embeddings of query entity $e_q$ and corpus entity $e_i$. The union of all retrieved entities forms the \textbf{Hit Entity Set} $E_H \subseteq E_I$. 

\paragraph{Relevant Chunk Retrieval and Scoring.}
Using the entity-to-chunks inverted index $M_{EC}$, we collect all chunks containing entities from $E_H$:
\[
C_H = \bigcup_{e \in E_H} M_{EC}(e)
\]
Each chunk $C_k$ is then scored based on two factors: (1) its semantic similarity to the query, and (2) the number of hit entities it contains:
\[
\text{score}_k = \phi_q \cdot \text{count}_k
\]
where $\phi_q = \phi(\text{emb}(C_k), \text{emb}(q))$ represents the query-chunk similarity, and $\text{count}_k$ is the number of hit entities referring to chunk $k$. By multiplying these factors, we ensure that chunks with both high semantic similarity and rich entity coverage receive the highest scores.

\paragraph{Context Selection and Assembly.}
We rank all chunks in $C_H$ by their scores and select the top-$K$ to construct the final context. To ensure coherence, selected chunks are reordered according to their original document position:
\[
C_{top} = \text{TopK}(C_H)
\]

The final context is assembled by merging the texts of the selected chunks, ensuring the total length stays within a predefined token limit:
\[
\text{Context} = \text{MergeTexts}(C_{top})
\]

This streamlined entity-guided retrieval pipeline enables SlimRAG to produce compact, highly relevant contexts without relying on complex graph structures or heavy traversal algorithms. By combining query decomposition with entity-level semantic matching and dual-factor scoring, the system achieves both precision in information retrieval and efficiency in context construction.

\section{Experiments}
\label{sec:experiments}
We conduct experiments to evaluate SlimRAG's effectiveness in improving retrieval-augmented generation, focusing on the challenging multi-hop question answering task. We aim to answer the following research questions:

\begin{itemize}
    \item \textbf{RQ1:} Does SlimRAG achieve competitive or superior retrieval accuracy and recall compared to existing retrieval-augmented generation (RAG) baselines?
    
    \item \textbf{RQ2:} Can SlimRAG reduce the overall token usage in index construction and mitigate retrieval redundancy (i.e., noisy or irrelevant context) while maintaining effectiveness?
    
    \item \textbf{RQ3:} How do the core components of SlimRAG individually contribute to downstream multi-hop QA performance?
\end{itemize}

\begin{table*}[t]
\centering
\caption{
Comparison on HotpotQA. \cmark{} = supported, \xmark{} = not supported.
"Index Type" indicates the retrieval structure; "Linking Mode" refers to how the query connects to context (e.g., graph traversal, chunk cluster, entity match); "Inc." = Incremental support. $\uparrow$ / $\downarrow$ indicate better directions.
}
\label{tab:main_results_hotpotqa}
\resizebox{\textwidth}{!}{%
\begin{tabular}{l c c c | c c c c c}
\toprule
\textbf{Method} & \textbf{Index Type} & \textbf{Linking Mode} & \textbf{Inc.} 
& \textbf{Acc (\%)~$\uparrow$} & \textbf{Recall (\%)~$\uparrow$} & \textbf{F1 (\%)~$\uparrow$} 
& \textbf{RITU~$\downarrow$} & \textbf{Index Time (s)~$\downarrow$} \\
\midrule
ZeroShot & - & - & \cmark & 35.467 & 42.407 & 38.705 & N/A & N/A \\
VanillaRAG & Flat  & Chunk Similarity & \cmark & 50.783 & 57.745 & 53.949 & N/A & N/A \\
RAPTOR & Tree & Chunk Clustering & \xmark & 55.321 & 62.424 & 58.661 & N/A & 2,934 \\
\midrule
GraphRAG & Graph & Entity Graph & \xmark & 33.063 & 42.691 & 37.377 & 67.363 & 3,512 \\
LightRAG (Low) & Graph & Entity Graph & \cmark & 34.144 & 41.811 & 37.631 & 56.513 & 26,754 \\
LightRAG (Global) & Graph & Entity Graph & \cmark & 25.581 & 33.297 & 28.837 & 56.513 & 26,754 \\
LightRAG (Hybrid) & Graph & Entity Graph & \cmark & 35.647 & 43.334 & 39.087 & 56.513 & 26,754 \\
\midrule
\rowcolor{lightgray}
\textbf{SlimRAG (Ours)} & \textbf{Index Table} & \textbf{Entity Matching} & \textbf{\cmark} 
& \textbf{57.414} & 59.325 & 58.355 & \textbf{16.31} & \textbf{2,164} \\
\bottomrule
\end{tabular}%
}
\end{table*}

\begin{figure}[t]
    \centering
    \includegraphics[width=1.0\linewidth]{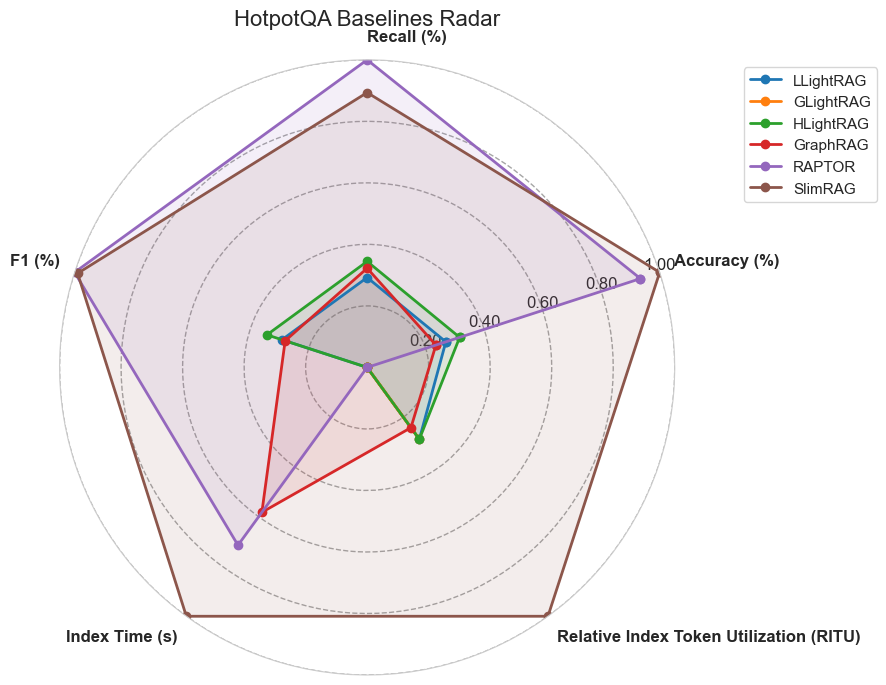}
    \caption{
\textbf{SlimRAG vs. Baselines on HotpotQA.}
Radar chart comparing SlimRAG and five baselines on accuracy, recall, F1, indexing time, and RITU (lower is better). Values are min-max normalized. SlimRAG shows strong overall performance with low indexing overhead.
}

    \label{fig:radar}
\end{figure}

\begin{figure}[t]
\centering
\includegraphics[width=1.0\linewidth]{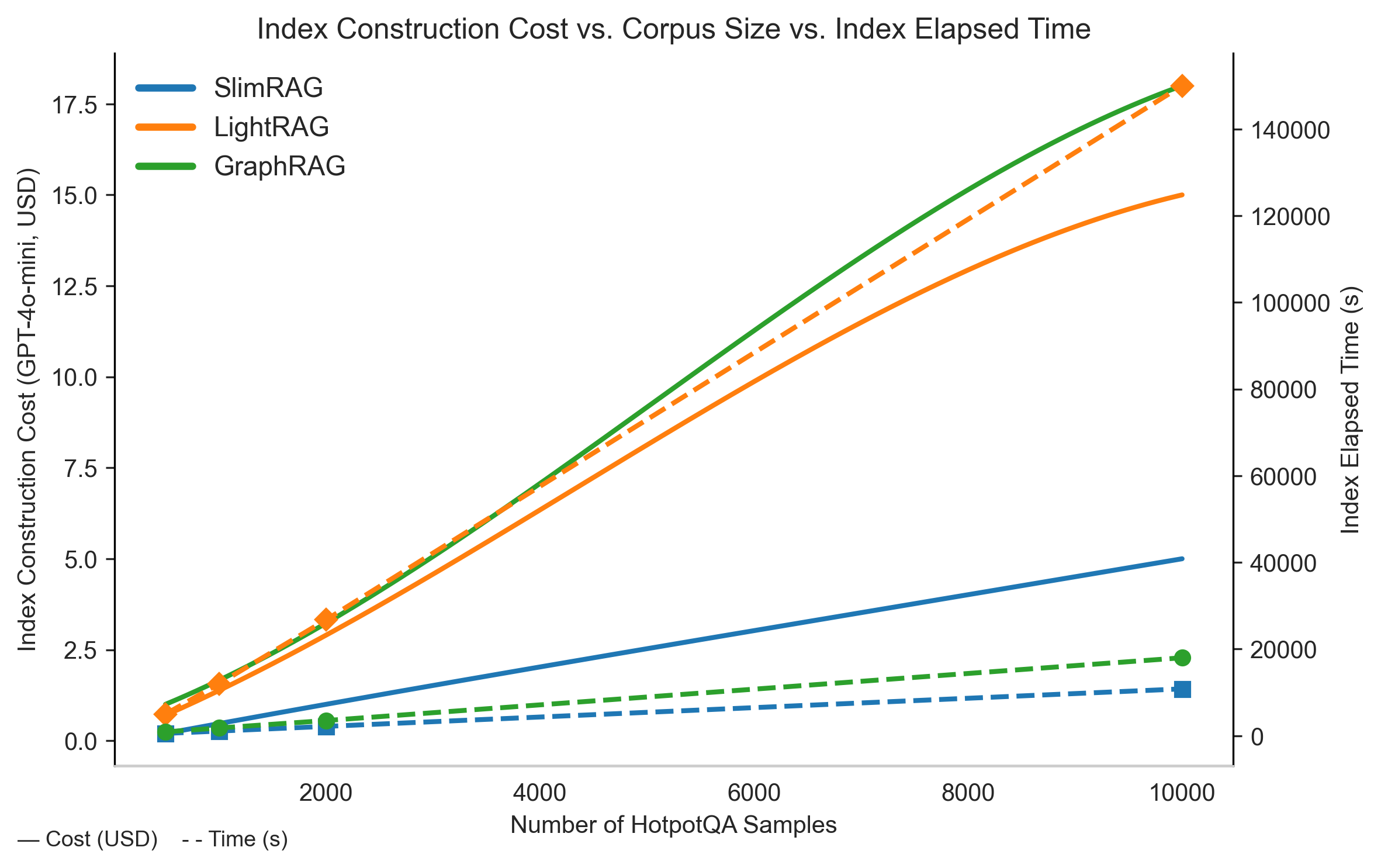}
\caption{
\textbf{Indexing Efficiency vs. Corpus Size.} 
Index construction cost (solid lines, USD) and elapsed time (dashed lines, seconds) for SlimRAG, LightRAG, and GraphRAG on HotpotQA. SlimRAG consistently incurs lower cost and latency as corpus size increases, highlighting its scalability advantage.
}

\label{fig:index_cost_vs_corpus_size}
\end{figure}

\subsection{Setup}
\label{ssec:setup}

\subsubsection{Dataset}
We evaluate SlimRAG on \textbf{HotpotQA}~\cite{yangHotpotQADatasetDiverse2018}, a widely used multi-hop QA benchmark that requires reasoning over multiple supporting documents. The knowledge corpus is the official Wikipedia dump released with HotpotQA, preprocessed into text chunks (such as individual sentences or short paragraphs) as described in our indexing pipeline.

\subsubsection{Baselines}
We compare \textbf{SlimRAG} against a set of representative retrieval-augmented generation (RAG) baselines. Our experiments on the HotpotQA dataset follow the system implementations and experimental setups described in~\cite{zhouIndepthAnalysisGraphbased2025}, ensuring consistency and fairness across all methods. Where applicable, we also reference or estimate retrieval statistics from original papers or widely recognized sources. Additionally, we report indicative token usage for both indexing and retrieval to assess the context efficiency of different methods.

\subsubsection{Evaluation Metrics}

We evaluate SlimRAG across a suite of retrieval and efficiency metrics. The central metric is \textbf{Relative Index Token Utilization (RITU)}, which captures the token-level compactness of the constructed index. Additional metrics assess retrieval effectiveness and system latency.

\begin{itemize}
    \item \textbf{Relative Index Token Utilization (RITU)}:  
    Defined as $\text{RITU} = \frac{\text{TUIC}}{\text{TCTC}}$, this dimensionless ratio quantifies the proportion of total corpus tokens actively used in index construction. A lower RITU indicates a more compact and efficient index.

    \item \textbf{Token Usage for Index Construction (TUIC)}:  
    The total number of tokens processed during indexing. For SlimRAG, this includes tokens used for entity extraction and association. For baselines, this includes all tokens embedded or used in graph construction.

    \item \textbf{Total Corpus Token Count (TCTC)}:  
    The number of tokens in the full raw corpus. Used as the denominator for RITU.

    \item \textbf{Index Time (sec)}:  
    The elapsed time required to build the index for first 2000 examples from HotpotQA subset. This measures real-world indexing latency under practical workload sizes.

    \item \textbf{Retrieval Accuracy (\%)}:  
    The proportion of retrieved chunks that match gold supporting facts, indicating retrieval precision.

    \item \textbf{Retrieval Recall (\%)}:  
    The proportion of gold supporting facts that are successfully retrieved, reflecting completeness.

    \item \textbf{Retrieval F1 Score (\%)}:  
    The harmonic mean of retrieval precision and recall, providing a balanced measure of overall retrieval quality.
\end{itemize}

\subsection{Main Results}
\label{ssec:main_results_hotpotqa}
We evaluate \textbf{SlimRAG} on HotpotQA by comparing it against a set of strong flat, tree-based, and graph-based RAG baselines. Table~\ref{tab:main_results_hotpotqa} reports results across retrieval structure, linking mechanism, and five evaluation metrics: accuracy, recall, F1, Relative Index Token Utilization (RITU), and index construction time. These are complemented by radar and growth analysis in Figures~\ref{fig:radar} and~\ref{fig:index_cost_vs_corpus_size}.

\subsubsection{Retrieval Performance: Accuracy, Recall, and F1}
Table~\ref{tab:main_results_hotpotqa} shows that \textbf{SlimRAG} achieves the highest retrieval accuracy (\textbf{57.41\%}) and F1 score (\textbf{58.36\%}) across all baselines. Its recall (\textbf{59.33\%}) is also highly competitive—slightly below RAPTOR (62.42\%), yet significantly outperforming flat baselines like VanillaRAG (57.75\%) and all graph-based variants of LightRAG (43.3\%).

Notably, despite being graph-free, SlimRAG closes or exceeds the performance gap with structure-heavy systems. This suggests that relying on clean entity-level signals can be more effective than over-engineered graph traversals, which often introduce spurious or irrelevant context links. The balanced strength across accuracy, recall, and F1 also indicates SlimRAG's ability to retrieve precise yet complete context for downstream reasoning.

\subsubsection{Retrieval Efficiency: RITU and Index Time}
SlimRAG delivers substantial efficiency gains in both token utilization and construction latency. It achieves the lowest \textbf{RITU} score (\textbf{16.31}), which is nearly \textbf{4×} smaller than GraphRAG (67.36) and LightRAG (56.51), indicating much higher index compactness. Additionally, SlimRAG’s index construction time for 2,000 samples is just \textbf{2,164 seconds}, markedly below LightRAG’s \textbf{26,754 seconds}, and faster than GraphRAG (3,512s) and RAPTOR (2,934s).

These results emphasize the practical benefits of SlimRAG’s design: by avoiding graph edge construction and redundant node processing, it builds an index that is not only lightweight but also faster to prepare. This efficiency is especially important in large-scale settings where index refresh or frequent domain adaptation is required.

\subsubsection{Retrieval Scalability: Cost and Latency under Growth}
Figure~\ref{fig:index_cost_vs_corpus_size} analyzes how indexing cost (in USD) and latency scale with the number of HotpotQA samples. SlimRAG exhibits the flattest growth in both cost and time. Its cost remains below \$0.30 per 1,000 samples, while LightRAG climbs steeply beyond \$1.50, and GraphRAG follows a similar trend. Latency grows similarly slowly for SlimRAG, whereas graph-based baselines scale poorly as corpus size increases.

These scalability trends highlight SlimRAG’s robustness under load. Unlike graph-based models that suffer exponential growth in both memory and time due to edge expansion and relational chaining, SlimRAG maintains near-linear scaling. This makes it particularly well-suited for production systems that need to operate over large or evolving knowledge corpora.

\subsection{Ablation Study}
\label{ssec:ablation}

To dissect SlimRAG’s core design contributions, we perform controlled ablation experiments focusing on two key modules: (1) \textbf{Coreference Resolution} during indexing, and (2) \textbf{Query Decomposition} during retrieval. As shown in Table~\ref{tab:ablation_results}, we evaluate all four combinations on HotpotQA to assess their individual and joint effects on retrieval performance.

\subsubsection{Effect of Coreference Resolution}
\label{sssec:ablation_coref}

Coreference resolution unifies co-referent mentions across the corpus, enabling accurate entity extraction. Ablating this component results in a steep drop in accuracy—from \textbf{57.41\%} to \textbf{40.98\%}—and recall—from \textbf{59.33\%} to \textbf{42.21\%}. This decline of over 16 points highlights its importance in resolving anaphoric references (e.g., pronouns or aliases) that string-matching alone cannot capture.

The absence of coreference handling causes SlimRAG to miss relevant chunks where entities are referenced indirectly, limiting multi-hop evidence recovery. As a result, both correctness and completeness deteriorate sharply. This also reveals the sensitivity of entity-based retrieval to linguistic variation and surface form inconsistency across documents.

\subsubsection{Effect of Query Decomposition}
\label{sssec:ablation_qd}

Query decomposition splits complex multi-hop questions into simpler sub-queries, improving targeted retrieval. Disabling this module reduces accuracy to \textbf{52.31\%} and recall to \textbf{52.98\%}, a relative drop of about 5 points. Although less dramatic than the coreference ablation, the performance loss confirms that decomposition enhances SlimRAG’s ability to cover all reasoning steps and retrieve distributed evidence more effectively.

This mechanism is particularly beneficial for questions requiring multiple logical hops or conditional sub-parts, where end-to-end matching underperforms.

\subsubsection{Joint Impact and Complementarity}

When both modules are removed, performance deteriorates further to \textbf{38.26\%} accuracy and \textbf{40.18\%} recall—the lowest observed across all configurations. This indicates a compounding effect:  
\begin{itemize}
    \item \textbf{Coreference resolution} ensures consistent entity anchoring across the corpus.
    \item \textbf{Query decomposition} expands coverage by aligning retrieval to all parts of the question.
\end{itemize}

The ablation findings confirm that both coreference resolution and query decomposition are indispensable to SlimRAG’s performance. Each module contributes distinctly—coreference resolution ensures accurate entity grounding, while query decomposition enables decomposable information access. Their synergy yields additive gains: neither component alone achieves full-system effectiveness, and their joint absence leads to severe degradation, demonstrating that linguistic normalization and structured querying jointly underpin SlimRAG's robust multi-hop retrieval.

\begin{figure*}[t]
  \centering
  \includegraphics[width=\linewidth]{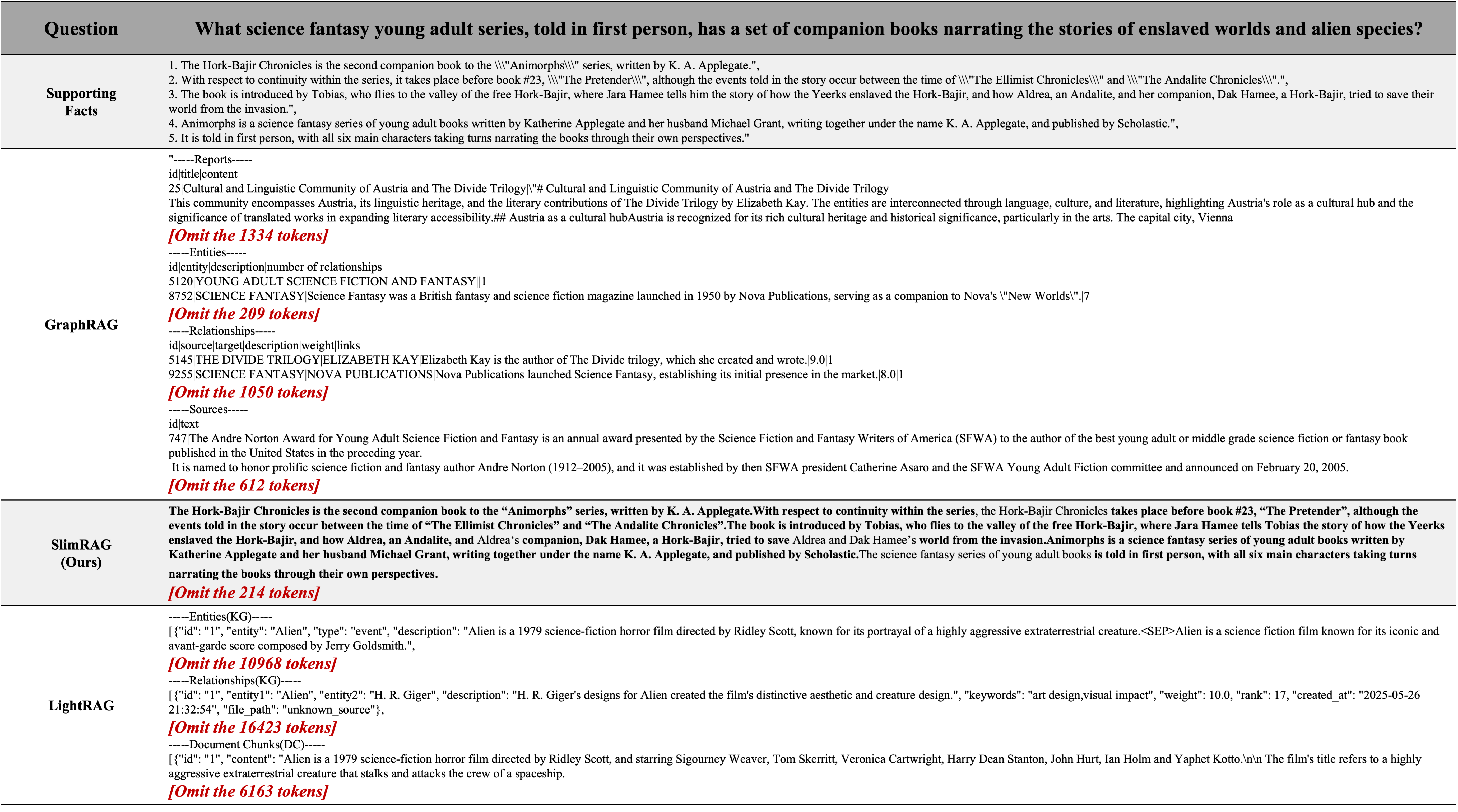}
  \caption{Case study on a sample multi‐hop HotpotQA question.  We show the top‐ranked context returned by LightRAG (34 137 tokens), GraphRAG (3 607 tokens), and SlimRAG (459 tokens).  SlimRAG’s output contains exactly the five gold‐supporting sentences required to answer the query, while the baselines retrieve large volumes of redundant or irrelevant text.  }
  \label{fig:case_study}
\end{figure*}

\subsection{Case Study}
To demonstrate the practical advantage of our decoupled indexing and retrieval strategy, we present a representative multi-hop HotpotQA example and compare the top-ranked contexts retrieved by three methods: LightRAG, GraphRAG, and SlimRAG. As shown in Figure~\ref{fig:case_study}, LightRAG retrieves over 34K tokens and GraphRAG over 3.6K tokens, with much of the content unrelated or tangential to the question. In contrast, SlimRAG retrieves only 459 tokens, precisely covering the five gold supporting facts required to answer the query.

This highlights two key benefits of SlimRAG: (1) substantial reduction in retrieval volume, which lowers computational overhead, and (2) focused evidence selection, which improves answer reliability. The case study illustrates how structure-heavy methods often introduce redundant content during retrieval, while our lightweight, entity-driven approach maintains high precision with minimal context.

\begin{table}[h!]
\centering
\caption{
Ablation results on HotpotQA. \cmark{} = used, \xmark{} = ablated. 
Coref = coreference resolution; Decomp = query decomposition.
}
\label{tab:ablation_results}
\resizebox{\columnwidth}{!}{%
\begin{tabular}{cc|ccc}
\toprule
\textbf{Coref} & \textbf{Decomp} & \textbf{Acc (\%)~$\uparrow$} & \textbf{Recall (\%)~$\uparrow$} & \textbf{F1 (\%)~$\uparrow$} \\
\midrule
\xmark & \xmark & 38.263 & 40.177 & 39.203 \\
\xmark & \cmark & 40.983 & 42.210 & 41.586 \\
\cmark & \xmark & 52.312 & 52.982 & 52.645 \\
\cmark & \cmark & \textbf{57.414} & \textbf{59.325} & \textbf{58.358} \\
\bottomrule
\end{tabular}
}
\end{table}

\section{Related Work}
\subsection{Retrieval-Augmented Generation: Foundations and Advances}
Retrieval-Augmented Generation (RAG) has emerged as a prominent paradigm to enhance the factuality, knowledge coverage, and updatability of large language models by enabling dynamic access to external corpora at inference time~\cite{lewisRetrievalaugmentedGenerationKnowledgeintensive2020,gao2023retrieval,huangSurveyRetrievalAugmentedText2024,fan2024survey}. The classic RAG pipeline consists of document segmentation (chunking), dense or hybrid retrieval, and conditional generation. Since its introduction, numerous variants have been proposed aiming to optimize recall, precision, efficiency, and reduce hallucination~\cite{shuster-etal-2021-retrieval-augmentation,ji2023survey}.

Recent research has intensified on query-aware and context-efficient retrieval, including methods that improve negative sampling~\cite{gao2022precise}, context compression~\cite{xu2023recomp,liu2023tcrallm} and grounding retrieval~\cite{qian2024grounding}. Alongside classic open-domain QA benchmarks (e.g., HotpotQA~\cite{yangHotpotQADatasetDiverse2018}, TriviaQA, Musique~\cite{trivedi2022musique}), RAG is increasingly central to long-context summarization~\cite{zhong2021qmsum,huang2021efficient}, multi-hop QA~\cite{ho2020constructing}, and dialogue~\cite{dinan2018wizard,shuster-etal-2021-retrieval-augmentation}. Multiple large-scale surveys~\cite{huangSurveyRetrievalAugmentedText2024,gao2023retrieval,fan2024survey} analyze the rapid evolution of RAG and emphasize challenges like retrieval redundancy~\cite{wuHowEasilyIrrelevant2024a,liEfficientDynamicClusteringBased2025} and task-specific relevance.

\subsection{Structured and Graph-based Retrieval in RAG}
To address document overlap, multi-hop reasoning, and robustness, many works have introduced structure-aware retrieval beyond flat chunking. Graph-based RAG methods build explicit knowledge graphs, entity graphs, or heterogeneous document-entity networks, enabling relational or multi-hop retrieval paths aligned with complex query semantics~\cite{edgeLocalGlobalGraph2024,guoLightRAGSimpleFast2024,chenSACKGExploitingLarge2024,somanBiomedicalKnowledgeGraphenhanced2023,sarmahHybridRAGIntegratingKnowledge2024,rezaeiAdaptiveKnowledgeGraphs2025,yangKGRankEnhancingLarge2024}. Techniques include entity extraction and linking, automatic discovery of relations, and retrieval of subgraphs for evidence assembly. Typical examples are GraphRAG~\cite{edgeLocalGlobalGraph2024}, LightRAG~\cite{guoLightRAGSimpleFast2024}, and KG-RAG-like systems~\cite{sanmartin2024kg, yangKGRankEnhancingLarge2024,li2024simpleeffectiverolesgraphs}.

However, the graph-based paradigm introduces new bottlenecks: (1) pipeline complexity (e.g., entity disambiguation, edge construction, coreference), (2) update and scalability challenges for large and dynamic corpora, and (3) retrieval redundancy due to spurious, tangential, or weak relational edges~\cite{xuGenerateonGraphTreatLLM2024,zhouIndepthAnalysisGraphbased2025}. Furthermore, recent analyses highlight that semantic similarity in embedding/graph space does not necessarily equate to semantic relevance for the query at hand~\cite{steckCosineSimilarityEmbeddingsReally2024,wuHowEasilyIrrelevant2024a}. This motivates efforts to separate structure from effective, query-centric retrieval.

\subsection{Retrieval Granularity, Compression, and Redundancy Mitigation}
Another active line of research investigates retrieval granularity and index compactness for RAG, balancing recall, efficiency, and overwhelm to downstream models~\cite{chenDenseRetrievalWhat2024,liEfficientDynamicClusteringBased2025,liangKAGBoostingLLMs2024}. Traditional approaches segment documents into fixed-size or sliding-window chunks, but recent work explores dynamic chunking, clustering~\cite{liEfficientDynamicClusteringBased2025,guoLightRAGSimpleFast2024}, and unsupervised community detection~\cite{traagLouvainLeidenGuaranteeing2019a} as offline redundancy reduction.

Adaptive, compression-based methods aim for query-aware context selection and token-level economy, such as passage filtering~\cite{wang2023learning}, selective augmentation~\cite{xu2023recomp}, and indexless/landmark strategies~\cite{luo2024bge,qian2024grounding}. Recent benchmarks~\cite{wuHowEasilyIrrelevant2024a,zhouIndepthAnalysisGraphbased2025} quantify token budget, supporting facts coverage, and index scalability. There remains a semantic gap between dense similarity (as used in most retrievers) and true, query-specific informativeness~\cite{chenDenseRetrievalWhat2024,steckCosineSimilarityEmbeddingsReally2024}.

\textbf{SlimRAG brings together these insights:} (1) eschewing expensive and brittle graph construction, (2) leveraging fine-grained, entity-aware inverted indexing, and (3) maximizing index compactness with novel metrics like Relative Index Token Utilization (RITU). Our work thus advances the goal of highly efficient, low-redundancy, and query-relevant retrieval-augmented generation.

\section{Conclusion}
\label{sec:conclusion}
In this work, we introduced \textbf{SlimRAG}: Retrieval without Graphs via Entity-Aware Context Selection. SlimRAG addresses a core limitation of existing RAG systems—the conflation of global semantic similarity with task-specific semantic relevance—by explicitly disentangling these notions across indexing and retrieval. It constructs compact indexes through coarse similarity-based clustering and performs fine-grained, entity-guided retrieval without relying on graph traversal or edge construction.
Experiments on multi-hop QA benchmarks show that SlimRAG improves answer accuracy (57.41\% vs. 55.32\%) while significantly reducing index redundancy, achieving a RITU of 16.31 compared to 56.51–67.36 for graph-based methods. We also introduce RITU as a general-purpose metric for evaluating token-level index compactness. These results highlight the efficiency and scalability of structure-free, modular retrieval, and suggest promising extensions to broader domains, larger corpora, and adaptive query decomposition strategies.

\bibliography{aaai25}

\section{Appendix}
\subsection{Foundational Data Structures and Functions}
This section establishes the formal groundwork for the SlinRAG algorithm detailed in Section Method. We introduce and define core data structures, sets, and functions crucial for understanding the operational stages of the algorithm.

Let $\mathcal{C} = \{D_1, D_2, \dots, D_M\}$ denote an unordered set of \textbf{Documents}, where each $D_k$ is a \textbf{Document}. Each document $D_k$ is itself an ordered sequence of sentences, $D_k = \langle s^{(k)}_1, s^{(k)}_2, \dots, s^{(k)}_{N_k} \rangle$, where $s^{(k)}_j$ denotes the $j$-th \textbf{Sentence} in document $D_k$, and $N_k$ is the total number of sentences in $D_k$. Each sentence $s^{(k)}_j$ is associated with a unique \textbf{Sentence ID} within its document, which, for simplicity, we can consider as its index $j$ in $D_k$.

Let $\mathcal{U}_E$ represent the universal set of all possible \textbf{Entities}. An entity $e \in \mathcal{U}_E$ is a distinct, identifiable object or concept within the text.

We define the following core components:
\begin{enumerate}
    \item \textbf{Entity Extraction Function}:\\
    $\text{ExtractEntities}: \mathcal{S} \rightarrow \mathcal{P}(\mathcal{U}_E)$\\
    This function takes a textual unit $\mathcal{S}$ (e.g., a sentence $s$ or a query $q$) as input and returns a set of entities identified within that unit. This function implicitly incorporates \textit{Coreference Resolution} to ensure that the returned entities are in their canonical forms. $\mathcal{P}(\mathcal{U}_E)$ denotes the power set of $\mathcal{U}_E$.

    \item \textbf{Index Entity Set ($E_I$)}:\\
    $E_I = \bigcup_{i=1}^{N} \text{ExtractEntities}(s_i)$\\
    $E_I$ is the set of all unique entities extracted and aggregated from all sentences in the corpus $\mathcal{C}$.

    \item \textbf{Index Entity-Sentence Mapping ($M_{IS}$)}:\\
    $M_{IS}: E_I \rightarrow \mathcal{P}(\{1, 2, \dots, N\})$\\
    This mapping associates each entity $e \in E_I$ with a set of Sentence IDs, representing all sentences in $\mathcal{C}$ that contain the entity $e$. Formally, for an entity $e \in E_I$, $M_{IS}(e) = \{i \mid e \in \text{ExtractEntities}(s_i)\}$.

    \item \textbf{Query ($q$)}: \\
    A natural language text string input by the user, representing their information need.

    \item \textbf{Query Entity Set ($E_Q$)}:\\
    $E_Q = \text{ExtractEntities}(q)$\\
    $E_Q$ is the set of entities extracted from the user's query $q$.

    \item \textbf{Embedding Function ($\text{emb}$)}:\\
    $\text{emb}: \mathcal{T} \rightarrow \mathbb{R}^d$\\
    This function maps a textual element $\mathcal{T}$ (which can be an entity $e$ or a sentence $s$) to a $d$-dimensional vector in a real-valued vector space $\mathbb{R}^d$.

    \item \textbf{Semantic Similarity Function ($\text{sim}$)}:\\
    $\text{sim}: \mathbb{R}^d \times \mathbb{R}^d \rightarrow \mathbb{R}$\\
    This function takes two $d$-dimensional vectors (e.g., $\text{emb}(x)$ and $\text{emb}(y)$) and returns a scalar value representing the semantic similarity between them (and thus, their corresponding original textual elements). This value is typically normalized, e.g., to the range $[0,1]$ or $[-1,1]$ (e.g., cosine similarity).

    \item \textbf{Context ($Ctx$)}:\\
    The final output of the algorithm for a given query $q$, consisting of an ordered set of $H$ sentences, denoted as $Ctx = \langle s'_1, s'_2, \dots, s'_H \rangle$. Each $s'_j \in \mathcal{C}$, and these sentences are arranged according to their original order in the corpus $\mathcal{C}$.
\end{enumerate}

\subsubsection{Experiments Details}
\begin{itemize}
    \item \textbf{Entity Extraction Function ($\text{ExtractEntities}$)}: The $\text{ExtractEntities}$ function is realized using OpenAI's \texttt{gpt-4o-mini} model. For any given textual unit (sentence $s$ or query $q$), \texttt{gpt-4o-mini} performs joint coreference resolution and multi-entity extraction. This process yields a set of unique, canonicalized entities, consistent with the function's formal definition.
    \item \textbf{Embedding Models}: All textual embeddings (for sentences $\text{emb}(s)$, queries $\text{emb}(q)$, and entities $\text{emb}(e)$) are generated using OpenAI's \texttt{text-embedding-3-small} model.
    \item \textbf{Semantic Similarity Function ($\text{sim}$)}: The semantic similarity ($\text{sim}$) between textual elements is calculated using cosine similarity on their respective \texttt{text-embedding-3-small} vectors.
    \item \textbf{Hyperparameters}: Key retrieval hyperparameters include $K=5$, which dictates the number of top similar entities retrieved during the semantic expansion of query entities, and $H=10$, which specifies the number of sentences in the final output context $Ctx$.
\end{itemize}

\end{document}